\documentclass{article}
\usepackage{ijcai18}

\usepackage{times}
\usepackage{color}
\usepackage{soul}
\usepackage{url}
\usepackage{graphicx}
\usepackage{subfigure}
\usepackage{float}
\usepackage[hidelinks]{hyperref}
\usepackage[utf8]{inputenc}
\usepackage[small]{caption}
\usepackage{amsmath} 
\usepackage{amssymb}  

\title{The Emergence of Complex Bodyguard Behavior Through Multi-Agent Reinforcement Learning}

\author{
Hassam Ullah Sheikh and
Ladislau B{\"o}l{\"o}ni
\\
Department of Computer Science\\
University of Central Florida\\
hassam.sheikh@knights.ucf.edu,
lboloni@eecs.ucf.edu
}

\begin{document}

\maketitle

\begin{abstract}
    In this paper we are considering a scenario where a team of robot bodyguards are providing physical protection to a VIP in a crowded public space. We show that the problem involves a complex mesh of interactions between the VIP and the robots, between the robots themselves and the robots and the bystanders respectively. We show how recently proposed multi-agent policy gradient reinforcement learning algorithms such as MADDPG can be successfully adapted to learn collaborative robot behaviors that provide protection to the VIP.
\end{abstract}

\section{Introduction}
\label{sec:Introduction}

One of the most complex challenges in autonomous robots concerns behaviors that require a complex interaction between robots and humans, as well as between the robots themselves. In particular, learning in these settings is challenging because the learner can not assume that the other robots/agents it is interacting with follow a stable strategy. To keep complexity manageable, most research work in this field is tested on simple problems, such as the cooperative communication or predator-prey, that involve the robot having to interact with only one or two types of other robots/agents.

In this paper we are considering a more complex problem  with a practical application. A VIP is navigating a crowded public area. We assume that the bystanders are purposefully moving from landmark to landmark, and use culture-specific crowd navigation protocols~\cite{Boloni-2013-RobotsInCrowds} on giving way or asserting their right of way. The goal of a team of robot bodyguards is to physically protect the VIP from assault by appropriately positioning themselves in relation to the current position of the VIP.


The goal of the bodyguards can be stated as a minimization of a metric of the  threat to the VIP, a problem made challenging due to the complexity of the environment. The bodyguard robot must adapt its position to the current position of the VIP. Second, the movement must also depend on the movement of the bystanders; if a bystander is heading towards the VIP, the robot needs to position itself to reduce the threat to the VIP. Finally, the robots need to coordinate between each other. This can be achieved either through central planning, explicit communication by exchanging messages or though implicit communication, by each robot taking actions based on a world view including the other robots.


While it is possible to hand-engineer robot bodyguard behaviors~\cite{bhatia:flairs-2016}, it is of interest whether such behavior can be learned through reinforcement learning.




\section{Problem Formulation}

The reward structure of the robotic bodyguard team problem can be modeled as a  cooperative Markov game where multiple agents are learning a policy to maximize their rewards. A multi-agent MDP can be defined as a combination of a state space $\mathcal{S}$, an action space $\mathcal{A}$, agents $\mathcal{A}_{1},\ldots,\mathcal{A}_{N}$, the transition probabilities that are defined as $\mathcal{T} = \mathcal{S}_{1} \times \mathcal{A}_{1} \times \ldots \times \mathcal{S}_{N}\times\mathcal{A}_{N}$ and the reward function for every agent that is defined as $r_{i}=\mathcal{R}\left(\mathcal{S}_{i}, \mathcal{A}_{i}\right)$.

For the sake of simplicity, we are assuming that all the bodyguards have an identical state and action space. We are considering a finite horizon problem where each episode is terminated after $T$ steps.

The environment that provides the rewards is a two-dimensional space, with the usual rules of physical movement, landmarks, line-of-sight and communication. We used the Multi-Agent Particle Environment MPE~\cite{mordatch:arxiv-2017} to perform our experiments. MPE is a two-dimensional physically simulated environment that runs in discrete time and continuous state and action space. The environment consists of N agents and M landmarks possessing physical attributes such as location, velocity and mass etc. MPE also allows the agents to communicate via verbal utterances over the communication channel embedded in the environment. The complete observation of the environment is given by
\[
o=\left[x_{1,\ldots N+M},c_{1,\ldots N}\right]\in O
\]
The state of each agent is the physical state of all the entities in the environment and verbal utterances of all the agents. Formally, the state of each agent is defined as
\[
s_{i}=\left[x_{j,\ldots N+M}, c_{k,\ldots N}\right]
\]
\noindent where $x_{j}$ is the observation of the entity $j$ from the
prespective of agent $i$ and $c_{k}$ is the verbal utterance of
the agent $k$.

\subsection{Reward Function}
\cite{bhatia:flairs-2016}~defined a metric that quantifies
the threat to the VIP from each crowd member $b_{i}$ at each timestep
$t$. We extended the metric to form a reward function defined as
\begin{equation}
\begin{aligned}
\mathcal{R}_{t} \left(\mathcal{B}, VIP, x_i \right)=&\displaystyle -1+\prod_{i=1}^{k}\left(1-TL\left( \textit{VIP},b_{i}\right)\right) \\&
+\mathcal{D}\left( \textit{VIP},x_{i}\right)+p
    \end{aligned}
\end{equation}
where
\[
TL\left(\textit{VIP},b_{i}\right)=\exp^{-A\left(Dist(\textit{VIP},b_{i})\right)/B}
\]
and
\[
\mathcal{D}\left(\textit{VIP},x_{i}\right)=\begin{cases}
0 & m \leq\left\Vert x_{i}-\textit{VIP}\right\Vert _{2}\leq d\\
-1 & \text{otherwise}\\
\\
\end{cases}
\]
\noindent where $m$ is the minimum distance the agents have to maintain at every timestep, $d$ is the safe distance and $p$ is a small penalty to the bodyguard for utterance.

\begin{figure}[ht]
\centering \includegraphics[width=0.7\columnwidth]{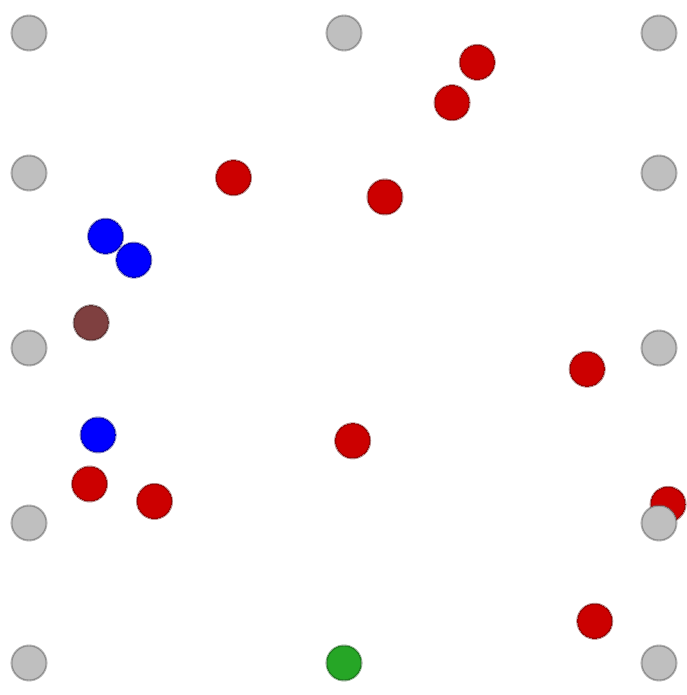}
\caption{Screenshot showing the emerging collaborative bodyguard behavior. The (brown) VIP aims to reach the green landmark. The (blue) bodyguard robots surround and separate it from the closest bystanders (red). The bystanders navigate among the gray landmarks.}
\label{fig:four_agent_binary_reward}
\end{figure}

\section{Methodology}
The bodyguards were trained using multi-agent deep deterministic policy gradients MADDPG~\cite{lowe:nips-2017}, a multi-agent extension of the DDPG algorithm described in~\cite{lillicrap:iclr-2015}. MADDPG allows agents to have individual policies, but trains them with a centralized Q-function. The gradient $\nabla J\left(\theta_{i}\right)$ of each policy is written as
\[
\nabla J\left(\theta_{i}\right)=\mathbb{E}\left[\nabla_{\theta_{i}}\log\pi_{i}\left(a_{i}|s_{i}\right)Q_{i}^{\pi}\left(s_{i},a_{1},\ldots,a_{N}\right)\right]
\]
\noindent where $Q_{i}^{\pi}\left(s_{i},a_{1},\ldots,a_{N}\right)$ is a centralized action-value function that takes the actions of all the agents in addition to the state of the agent to estimate the Q-value for agent $i$. The primary motivation behind MADDPG is that knowing all the actions of other agents makes the environment stationary, even though their policy changes.

\section{Experiments}

The experiments were conducted in an environment created in MPE, simulating a crowded mall with 12 landmarks representing stores. The VIP is following its own policy of moving towards a goal landmark. 10 bystanders are following their own policy that involve sequentially visiting randomly chosen landmarks. 3 bodyguards were deployed to protect the VIP from physical assault.

For training, the bodyguard agents were trained over 10,000 episodes, each episode being limited to 25 steps using DDPG and MADDPG. After training, performance was measures both empirically (by means of observing the movement of the bodyguards) and by recording the cummulative threat over the episode as defined in~\cite{bhatia:flairs-2016}. An example of the achieved behavior is described in Figure~\ref{fig:four_agent_binary_reward}. As the screenshot shows, the robots learned to surround and move with the VIP, concentrating on the sides where bystanders are closer. Empirical evaluation had shown better performance with MADDPG compared to DDPG. 




\section{Conclusion}

In this paper, we outlined a technique for training a multi robot team of bodyguards that is collaborating towards a common goal of providing
physical security to a VIP moving in a crowded environment. We trained the robot behaviors using MADDPG RL, and we observed the emergence of recognizable collaborative bodyguard behavior without any explicit instruction on how to provide security.

\bibliographystyle{named}
\bibliography{ijcai18}

\begin{thebibliography}{}

\bibitem[\protect\citeauthoryear{Bhatia \bgroup \em et al.\egroup
  }{2016}]{bhatia:flairs-2016}
T.S. Bhatia, G.~Solmaz, D.~Turgut, and L.~B{\"o}l{\"o}ni.
\newblock Controlling the movement of robotic bodyguards for maximal physical
  protection.
\newblock In {\em Proc of the 29th International FLAIRS Conference}, pages
  380--385, May 2016.

\bibitem[\protect\citeauthoryear{B{\"o}l{\"o}ni \bgroup \em et al.\egroup
  }{2013}]{Boloni-2013-RobotsInCrowds}
Ladislau B{\"o}l{\"o}ni, S~Khan, and Saad Arif.
\newblock Robots in crowds-being useful while staying out of trouble.
\newblock In {\em Proc. of Intelligent Robotic Systems Workshop (IRS-2013) at
  AAAI}, pages 2--7, 2013.

\bibitem[\protect\citeauthoryear{Lillicrap \bgroup \em et al.\egroup
  }{2015}]{lillicrap:iclr-2015}
Timothy~P. Lillicrap, Jonathan~J. Hunt, Alexander Pritzel, Nicolas Heess, Tom
  Erez, Yuval Tassa, David Silver, and Daan Wierstra.
\newblock Continuous control with deep reinforcement learning.
\newblock In {\em Proceedings of the International Conference on Learning
  Representations (ICLR)}, 2015.

\bibitem[\protect\citeauthoryear{Lowe \bgroup \em et al.\egroup
  }{2017}]{lowe:nips-2017}
Ryan Lowe, Yi~Wu, Aviv Tamar, Jean Harb, Pieter Abbeel, and Igor Mordatch.
\newblock Multi-agent actor-critic for mixed cooperative-competitive
  environments.
\newblock {\em Neural Information Processing Systems (NIPS)}, 2017.

\bibitem[\protect\citeauthoryear{Mordatch and
  Abbeel}{2017}]{mordatch:arxiv-2017}
Igor Mordatch and Pieter Abbeel.
\newblock Emergence of grounded compositional language in multi-agent
  populations.
\newblock {\em arXiv preprint arXiv:1703.04908}, 2017.

\end{thebibliography}

\end{document}